\documentclass[num-refs]{wiley-article}
\pdfoutput=1
\usepackage{siunitx}

\usepackage{graphicx}
\usepackage{color}
\usepackage{comment}
\usepackage{xspace}
\usepackage{float}

\newcommand{\ie}{\emph{i.e.,}\xspace}
\newcommand{\eg}{\emph{e.g.,}\xspace}

\newcommand{\etal}{\emph{et al.}\xspace}

\title{Machine Learning Software Engineering in Practice: An Industrial Case Study}

\author[1\authfn{1}]{Md Saidur Rahman}
\author[1\authfn{1}]{Emilio Rivera}
\author[1\authfn{1}]{Foutse Khomh}
\author[1\authfn{1}]{Yann-Gaël Guéhéneuc}
\author[2\authfn{2}]{Bernd Lehnert}

\affil[1]{DGIGL, Polytechnique Montreal, Montreal, QC, H3T 1J4, Canada}
\affil[2]{SAP Canada Inc, Montreal, QC,  H3C 2M1, Canada}

\corraddress{Md Saidur Rahman, Polytechnique Montreal, Montreal, QC, H3T 1J4, Canada}
\corremail{saidur.rahman@polymtl.ca}

\presentadd[\authfn{1}]{DGIGL, Polytechnique Montreal, Montreal, QC, H3T 1J4, Canada}

\runningauthor{Rahman et al.}

\begin{document}

\maketitle

\begin{abstract}
SAP is the market leader in enterprise software offering an end-to-end suite of applications and services to enable their customers worldwide to operate their business. Especially, retail customers of SAP deal with millions of sales transactions for their day-to-day business. 
Transactions are created during retail sales at the point of sale (POS) terminals and then sent to some central servers for validations and other business operations. 
A considerable proportion of the retail transactions may have inconsistencies due to many technical and human errors. SAP provides an automated process for error detection but still requires a manual process by dedicated employees using workbench software for correction. However, manual corrections of these errors are time-consuming, labor-intensive, and may lead to further errors due to incorrect modifications. This is not only a performance overhead on the customers' business workflow but it also incurs high operational costs. Thus, automated detection and correction of transaction errors are very important regarding their potential business values and the improvement in the business workflow. In this paper, we present an industrial case study where we apply machine learning (ML) to automatically detect transaction errors and propose corrections.
We identify and discuss the challenges that we faced during this collaborative research and development project, from three distinct perspectives: Software Engineering, Machine Learning, and industry-academia collaboration. 
We report on our experience and insights from the project with guidelines for the identified challenges. We believe that our findings and recommendations can help researchers and practitioners embarking into similar endeavors.

\end{abstract}

\section{Introduction}
Artificial Intelligence (AI) and Machine Learning (ML) have shown promising prospects for intelligent automation of diverse aspects of business and everyday life \cite{Foutse_Redhat_2018}. 
The rapid development of machine learning algorithms and tools, and easier access to available frameworks and infrastructures have greatly fueled this era of the development of ML-based software solutions for real-world problems. 
Software companies, big or small, are striving to adopt ML in software applications, in order to improve their products and services. 
However, the promising potentials of machine learning accompany multifaceted challenges to the traditional software development processes and practices \cite{Schelter_IEEE_bulletin_2018, Foutse_Redhat_2018}. 

The development of ML-based software applications may add challenges to all phases of the  development life cycle \cite{Amershi_ICSE_2019}. 
The requirements for the ML models are expected to be dynamic in nature; to adapt to the rapidly evolving user requirements and business cases. ML-based applications are data-driven. Thus, efficient pipelines and infrastructures are required for data-streaming; i.e., data acquisition, data storage, preprocessing and feature extraction, and ingestion by the ML models. Also, model building and deployment have different constraints regarding the type of problem, data, and the target environments. Software engineering for machine learning applications has distinct characteristics that render most traditional software engineering methodologies and practices inadequate~\cite{Schelter_IEEE_bulletin_2018}. The design of ML applications needs to be flexible to accommodate the rapidly evolving ML components. In addition, the testing of ML applications differs significantly from traditional software testing. So, new guidelines are needed to help ML developers cope with these challenges. %
Additional challenges are likely to surface when the ML applications are developed in collaboration of multiple teams with diverse backgrounds and expertise~\cite{Garousi_IST_2017,Sandberg_ICSE_SEIP_2017}. 
In addition, there is an utmost need for guidelines for best practices in collaborative research and development between industry and academia in the context of software engineering for ML applications.     

In this paper, we present an industrial case study to share our experience and insights from a research project on machine learning conducted in collaboration between Polytechnique Montreal and SAP Inc. 
We first present our approach to developing ML-based components for automatic detection and correction of transaction errors. 
Here, we apply machine learning on retail transaction data to detect and correct transaction errors. 
We follow an agile methodology to develop ML-based solution. 
We then identify the challenges in each step of our software development process. 
We analyze and map these challenges along three distinct perspectives: (1) software engineering, (2) machine learning, and (3) industry-academia collaboration. From these perspectives, we explore the relationships and dependencies among the challenges. This helps us to have better insights into the challenges in software engineering for machine learning, especially in a collaborative research and development context. 

Our key contributions are summarized bellow: 
\vspace{-0.1cm}
\begin{itemize}
    \item We have identified important challenges in software engineering for machine learning in the context of industry-academia collaboration. Our insights will help developers building ML applications. 
    \item Based on our experience from the industrial case study, we propose guidelines for researchers and practitioners for best practices in the development of ML applications. 
\end{itemize}
\vspace{-0.05cm}
The rest of the paper is organized as follows: 
Section \ref{lblBackground} outlines the relevant  background and concepts. 
Section \ref{lblCaseStudy} presents our approach to apply machine learning for the detection and correction of transaction errors. We share important lessons that we learned from this case study in Section \ref{lblLessons}. We describe our insights from the perspective of industry-academia collaborative research and development for machine learning applications. 
Section \ref{lblRelatedWorks} discusses the related works.
Finally, Section \ref{lblConclusion} concludes the paper. 
\vspace{-0.15cm}
\section{Background}\label{lblBackground} 
To present the basic concepts of transaction errors and their corrections, we briefly discuss the structure and organization of the transaction data, the error characteristics and the correction procedure in the following subsections.
\vspace{-0.25cm}
\subsection{Retail Transaction} \label{lblTransAndErrors}
A retail transaction generally created in POS terminals represents information about a business activity such as the sale or return of item(s) and the payments. 
In this study, by \textit{transaction}, we refer to a set of records (table rows) in the transaction database that is related to a single order from a retail customer. 
Component rows of a particular transaction can be identified by the keys in database tables. Transactions contain the details of the product items, quantity, price, discounts, taxes and various other informations required for the business functions. 
Transactions also contain meta-data as the transaction header. 
\vspace{-0.2cm}
\subsection{Transaction Data Structure}
\begin{figure*}
    	\begin{center}
        \includegraphics[scale=0.5]{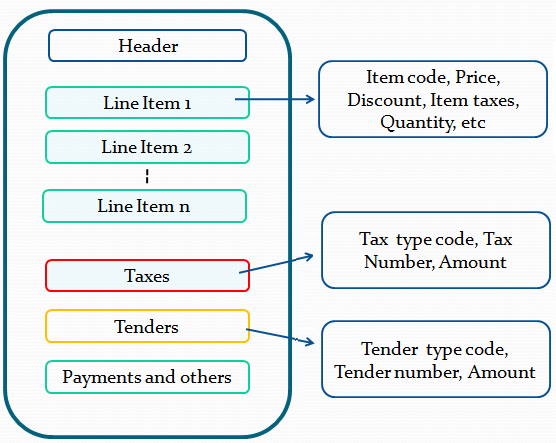}
        \caption{Transaction data structure}\label{figTransaction}
        \end{center}
\end{figure*}

A transaction (as shown in Fig. \ref{figTransaction}) consists of different components organized in a hierarchy. 
The header contains metadata about the transaction itself. 
The Components represent either item-level information (\eg price, product code, quantity, item-specific discounts) or the transaction-level information(\eg  transaction type, taxes, discounts, payments). 
In hierarchical (tree-like) organization, the transaction header is the root and other components are children. Each component can have zero or more sub-component(s). For example, an item can have none or several discounts applied on it. Attribute values are at the leaf nodes in the hierarchy. For each record (row) in the transaction table, the field (\textit{record qualifier})  indicates the type of the record (\eg header, discount, tax etc.). The \textit{parent} field defines the parent row of the current row. 
The relationships among the transaction components are governed by well-defined business rules. 
%

\subsection{Transaction Data Organization}
The transaction data are primarily contained in two (2) flattened tables: a Transaction Log (TLOG) table representing the details of the transactions and a Process Log (PLOG) table logging the details of the changes to the transactions (in TLOG). 
%
A tuple of four (4) fields including the transaction index, transaction date, store number, and the time stamp uniquely identify each transaction in TLOG. 
The same fields as in the TLOG table can be used as the key in the PLOG table for the identification of individual transactions. 
These key fields are used for mapping of transactions between TLOG and PLOG tables. 
\vspace{-0.25cm}
\subsection{Transaction Errors and Corrections} \label{lblErrors}

Transactions created at POS terminals are sent to the SAP's validation system. 
The validation involves a series of tests that can either succeed or fail. 
If succeeds, a transaction moves to the next task until the validation is complete. 
In case of failure, the cause of the failure (error) is logged in the PLOG table. 
These transactions are the candidates for manual corrections.  
%
%
PLOG table keeps track of changes in transaction fields by logging the changes with three (3) fields: field name, old value, and new value. 
Here, the error status can be used to detect the error while the changes to fields can identify the types of errors. 
%
%
Manual correction of transactions is done by operators using SAP workbench. 
The operators may need to update (add, remove or modify) the attributes in transaction data to correct errors.  
We leverage this correction logs and the transaction data to train machine learning models for error detection and correction. 

\vspace{-0.2cm}
\section{The Case Study} \label{lblCaseStudy}
This section presents our case study. Here, we describe our approach to develop the ML component 
for automatic detection and correction of errors in retail transactions. 
\vspace{-0.2cm}
\subsection{The Problem} \label{lblProblem} 
Transactions are the primary business data in retail industries and the transactions need to be correct and consistent. However, as mentioned earlier, due to many technical and human errors, transactions can have inconsistencies. These errors need to be corrected for related business functionalities (\eg audits) and data integrity. Currently, due to the lack of an automated or semi-automated tool, these errors are corrected manually by dedicated employees of customer companies of SAP. Manual correction incurs high costs and it is a serious bottleneck to the business operations. Thus, the problem we address in this case study is the development of ML-based components 
for automatic detection and correction of transaction errors.  
We divide the problem into two precise sub-problems: (1) \textit{detection or classification of the transaction errors} and (2) \textit{recommending appropriate actions (operations and values) to correct the transactions}. 
Here, our objective is to build machine learning models to detect (classify) and to correct (predict actions and the correct values) errors in new transactions and to integrate the new components into the existing SAP solutions. 


\vspace{-0.15cm}
\subsection{Overall Solution Architecture}
\begin{figure*}
	\begin{center}
		\includegraphics[scale=0.5]{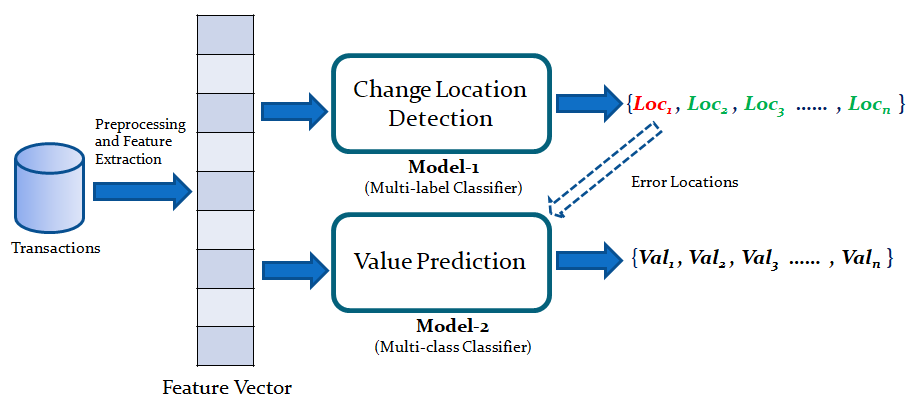}
		\vspace{-0.25cm}
		\caption{Error detection and correction system architecture}\label{figDiagram}
		\vspace{-0.55cm}
	\end{center}
\end{figure*}

 As shown in Fig. \ref{figDiagram}, we process transactions data from the database and extract features. Then, we apply machine learning on the features to build ML-models for the components for detection and correction of errors. 
 We take a two-step approach. Our first model (Model 1) is a multi-label classifier that detects the types of errors in the erroneous input transaction. Each label in the classification corresponds to a specific error type (and location) of the error. Once we have the errors detected, we apply our second model (Model 2) to predict values for the correction of the error. 
 Model 2 is a multi-class classifier that predicts the correction value from the set of possible values for a transaction field. 
\vspace{-0.2cm}
\subsection{Data Acquisition}
As mentioned earlier, transaction data is stored in a large flat table (TLOG). Another table (PLOG) captures all changes to transactions that are done either manually or automatically. Although there are millions of transactions, we are particularly interested in the erroneous transactions that were successfully corrected. Based on the error (task) status of the transactions from PLOG, we select corresponding erroneous transactions from the TLOG table. 
Again, the TLOG table contains the updated state of the transactions. Thus, once corrected, only the corrected version of an erroneous transaction is available in the TLOG. 
However, for ML algorithms, we need the erroneous state of the transactions from the TLOG. We apply reverse engineering on transactions in TLOG based on the change-logs in the PLOG to extract the erroneous version of the transactions. 
The transactions from the TLOG and the change logs in PLOG are the primary sources of data for our analysis. 
\vspace{-0.25cm}
\subsection{Preprocessing Transaction Data}
Transaction data is passed through some transformations at SAP customers' end to remove sensitive business and personal information. 
Not all of the transactions with correction logs qualify for our analysis. 
We filtered out transactions 
that are missing necessary attributes or contain inconsistent field values, or that are outliers (with an excessive number of product items). 
\vspace{-0.1cm}
\subsection{Feature Extraction}
To extract features, we programmatically navigate the structures of the selected transactions to access transaction fields to extract base features. 
We also compute derived features from the extracted base features. 
Each feature vector is composed of features representing both transaction-level and item-level information. 
As transactions are variable in size regarding the number of items, we consider transactions only with maximum 20 retail items to avoid making the feature too sparse. This threshold is based on our manual analysis of the distribution of the number of retail items per  transaction. 
We define a fixed-length feature vector that contains information for a maximum of 20 retail items. 
We process all the qualifying transactions and store the features in Hierarchical Data Format (HDF) file for efficient processing. 
Our selection of features are based on the knowledge from domain experts and our understanding of the relationships among the transaction components. We iteratively refined our choice of the transaction fields as features for ML algorithms.  
\vspace{-0.2cm}
\subsection{Error Detection Approach}
By error detection, we refer to the identification of the types and locations of errors.  
A transaction can have multiple errors. 
Thus, we model our error detection problem as a multi-label classification problem. 
Here, we label the feature vector for each transaction with a binary vector of length equal to the number of possible error classes. The label vector contains 1s in places where there is an error in the corresponding transaction component and 0s otherwise. We identify errors by examining the manual changes in the process log (PLOG) table. 
For example, if we detect a change in a tender type code, we infer that there was an erroneous tender type code. Thus, we set the bit in the label vector corresponding to the tender type code to 1. 
Once we have a labeled dataset, we build machine learning models to perform multi-label classification for error detection. We divide our dataset into three subsets for training, testing, and validation. 
\vspace{-0.1cm}
\subsubsection{The Dataset}
For error detection, we need transaction data with correction examples. So, we select transactions that were erroneous and correction histories in PLOG. 
We exclude erroneous transactions without corrections. 
Then, we extract features from the erroneous version of the corrected transactions. 
We also add samples from transactions without any errors in our dataset. These transactions qualify all the selection criteria (\eg no of items) except they are not erroneous. We split the features into balanced data sets for training, testing, and validation.  

\subsubsection{Multi-label Classifier}
We apply Random Forest algorithm for multi-label classification. Random Forest algorithm can be applied for both classification and regression problems and can handle the over-fitting problem of Decision Tree algorithm. Random Forest algorithm is an ensemble learning method that constructs a multitude of decision trees. The classification and prediction decisions are based on the combination of output of the constructed decision trees. In addition, we also apply Decision Tree, and AdaBoost Classifier to detect errors in transactions. However, RandomForest outperformed both of these algorithms. We also apply neural networks (NN), specifically Multilayer Perceptron (MLP), for error classification. However, because of our limited number (less than 100 to a few hundred in most cases) of error correction samples, MLP could not be trained properly for most of the error types. 
Finally, we opted for Random Forest which achieved the best performance. 

\vspace{-0.25cm}
\subsection{Error Correction Approach}
By error correction, we refer to predicting the correct values for the erroneous transaction components. 
We model the value prediction problem as a multi-class classification problem. 
The predicted class refers to a value from the possible set of values. Each model is trained for a distinct type of error.
The features are extracted from the erroneous transactions in TLOG and the corresponding logs in the PLOG. 
The label is determined by examining the new value and old value of the field related to the transaction error. The new value of a changed field of an erroneous transaction is set as the target label for the algorithm to predict. For the categorical field values (e.g., Tender Type Code), we need to predict the correct value from the set of alternative values for a given field. In such case, the target label is a one-hot encoded vector of possible values for the associated field. 
\vspace{-0.15cm}
\subsubsection{The Dataset}
We build our error correction model based on the transactions with corrections available in the transaction database. We split the features into balanced data sets for training, testing, and validation.

\vspace{-0.15cm}
\subsubsection{Multi-class Classifier}
We formulate the error correction problem as a multi-class classification problem. Here, the classifier picks one of the values from all possible alternative values as the correct value. The label vector for each data sample represents a binary vector with '1' in the index of the correct value and 0's in all other places. 
We apply the Logistic Regression (one versus all) algorithm for the value-prediction problem. We also applied neural networks (MLP) to predict values for error correction. 
As we have mentioned earlier, we have limited correction data samples for most of the error types. Consequently, NN models performed low compared to simple Logistic Regression. Limited data size and class-imbalance issues are likely to blame for the poor performance of NNs.

\vspace{-0.25cm}
\subsection{Metrics}

To measure the performance of the models, we measure the precision, recall, accuracy score and Jaccard similarity score of the models \cite{PrecisionRecallAccuracy,JaccardSimilarity}. Here, \textit{precision} refers to the ratio of the number of correctly predicted or classified cases to the total number of cases. We measure \textit{recall} as to the ratio of the total number of correctly predicted or classified cases to the total number of true cases. \textit{Accuracy}, on the other hand, refers to the ratio of correct prediction to the total number of prediction. 
We also measure \textit{Jaccard Similarity} Score or Jaccard coefficient which is the ratio of the cardinalities of the intersection and the union of two sets of labels (predicted and the true set). For binary classification, this metric is equivalent to the accuracy score while they differ for multi-label classification. 
\subsection{Model Performance} \label{lblPerformance}
This section presents the performance of the models for the detection and correction of errors in retail transactions. We present the performance of the models that detect "Tender Type code" errors.  
\subsubsection{Error Detection}
\begin{figure*}[h]
     	\begin{center}
         \includegraphics[scale=0.8]{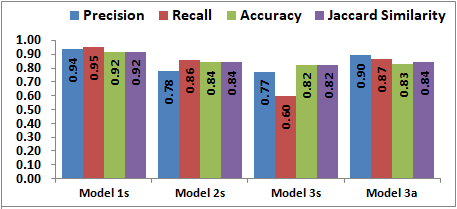}
         \vspace{-0.25cm}
         \caption{Error detection performance}\label{figDetection}
         \end{center}
 \end{figure*}

Fig. \ref{figDetection} presents the error detection performances of four different model configurations regarding error types. Here, Model1s, Model2s, and Model3s are to predict errors only in the first, second and third tender type code respectively. Model3a is trained to predict all of the first three tender type code errors. We present the error detection performance based on the Random Forest algorithm. 

Here, the first group of bars in Fig. \ref{figDetection} (Model1s) represents the error detection performance when only errors in the first tender type code were considered. Here, we observe that the detection accuracies of the model are above 90\%. However, for the second (Model 2s) and third (Model 3s) models, the detection performance drops. This declination in performance might be related to the comparatively lower number of training samples for each category. Again, when we include all the first three tender type codes in the detection model (Model 3a), we see a slight increase in the detection performance (above 80\%). We observe that the size of the training dataset influences the training of the models and 
their prediction accuracies. 

\subsubsection{Value Prediction}
\begin{figure*}
     	\begin{center}
         \includegraphics[scale=0.95]{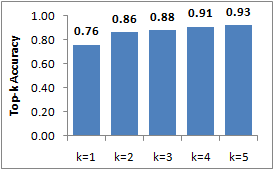}
         \caption{Top-k recommendation performance}\label{figCorrection-topk}
         \end{center}
 \end{figure*}

Fig. \ref{figCorrection-topk} presents the accuracy of top-k recommendation of correction values. This result shows the accuracy of predictions (i.e., recommendations) when a correct value exists in the top-k of the ranked list of recommended values.  
Fig.~\ref{figCorrection-topk}, shows performance results obtained by applying the Logistic Regression Cross-validation (one-vs-rest) algorithm for multi-class classification. 
%
%
 %
Here, we see that the value prediction accuracy of the model is 76\% if we consider single predicted value with the highest probability (\ie k=1). 
However, we also evaluate the prediction accuracy for top-k (k = [1, 5]) of the model. We see that the accuracy of top-5 prediction containing the expected value is 93\%. 

\vspace{-0.2cm}
\subsection{Agile Approach for Research and Development}
We adopted the agile methodology (SCRUM \cite{Schwaber_SCRUM_2002}) considering the incremental and iterative nature of the problem. SCRUM is a widely used agile methodology in the industry as it facilitates close collaboration and interactions between team members. SCRUM accelerates software development with faster delivery cycles while offering flexibility \cite{Sandberg_ICSE_SEIP_2017}. In our industry-academia collaborative team, academic members have expertise in both software engineering and machine learning. Members from industry partner (SAP) have expertise on design, implementation, and maintenance of SAP software solutions, fundamental understanding of machine learning, and the detail domain knowledge of the SAP technology and business models. Here, the `product owner' was from the industry partner and the `SCRUM master' was from the academic partner. The length of each `sprint' was 30 days with `daily stand' to update on changes towards the sprint goals. There were weekly meetings to review and to sync progress while there were biweekly meetings for all members from academia and industry to review the progress on the sprint milestones and deliverables. Each sprint ended with a `sprint demo' and review of any `retrospective' to address backlogs. The whole project spanned over six sprints. 
\section{Lessons Learned} \label{lblLessons}
Machine learning applications have some distinct characteristics compared to traditional software applications. Thus, software developers and practitioners should be aware of a number of challenges or risk factors regarding Machine Learning applications \cite{Sculley_TechDebt_2014}. To shed light on the common challenges in ML application development, we describe our experience from the case study along three different perspectives: software engineering (S), machine learning (M) and the industry-academia collaboration (C) for research. As in Fig. \ref{figDimensions}, we represent different aspects regarding each of the three dimensions of challenges. We explore their relationships to have insights into the software engineering principles and practices for machine learning applications. We focus on each dimension and identify different important challenges that need to be considered in an ML application development process. 
It is to be noted that each of the dimensions in Fig. \ref{figDimensions} linearly lists the challenges associated with different development phases or contexts which are iterative in nature. 
\subsection{Software Engineering Perspectives} \label{secSE}
Machine learning applications, like other software systems, need a well-defined software engineering process for its development and maintenance. However, given the distinct characteristics of the ML applications, the phases of the software engineering process may need adjustments to accommodate the ML specific requirements. We discuss different phases of software development life cycle (SDLC) for ML applications in as follows:    
\begin{figure*}[ht]
	\begin{center}
		\includegraphics[scale=0.6]{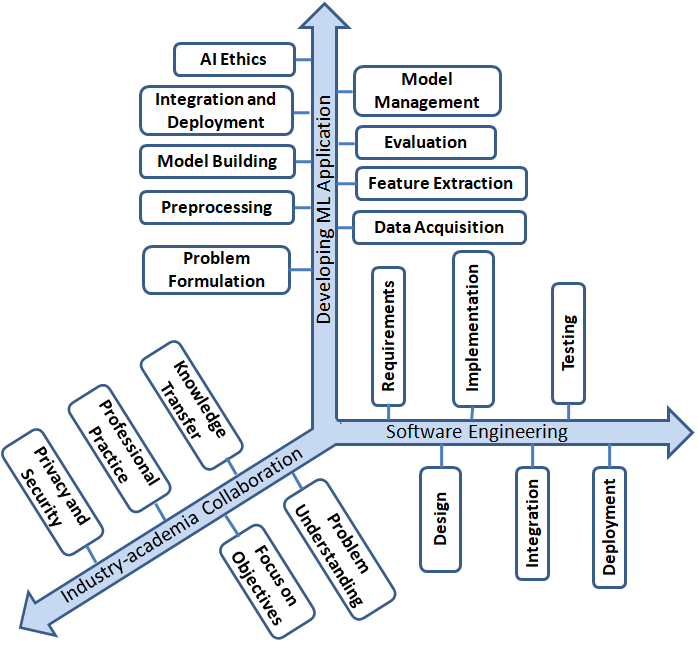}
		\caption{Dimensions of Challenges in Industry-academia Collaborative ML Research Project}\label{figDimensions}
		\vspace{-0.7cm}
	\end{center}
\end{figure*}

\begin{itemize}
	\item [\textbf{S1:}] \textbf{\textit{Requirements Engineering}:}
	Poor quality of requirements can lead to many issues in the successive phases of the software development \cite{Firesmith_JOT_2007}. 
	Requirement engineering for ML applications involves both ML specific and traditional requirements engineering activities such as feasibility analysis, requirements gathering, requirement specification and validation. 
	As the requirements in ML applications may change frequently, requirements specification for ML applications is a challenging task \cite{Attarha_ICIS_2011}. 
	%
	%
	In our project, we gathered requirements from the discussions and demonstration by the domain experts. We also observed the transaction data structures, example error types, and their correction procedures. 
	These helped us to 
	specify functional and non-functional requirements. 
	We iteratively refined our requirements specification based on the feedback from the domain experts. 
	
	In our project, the members from the academic partner did not have direct interactions with the end-users (SAP client). 
	The requirements information was gathered and shared by the industry partner. We observed that direct communication with the end-user and the on-site observation of error correction scenarios could be more useful in faster requirement specification and problem formulation. Interpretation differences of the requirements  due to indirect communications may cause delayed identification of some requirements. Such delayed identification or changes in requirements can be costly. 
	
	\item [\textbf{S2:}] \textbf{\textit{Design}:} 
	Software design specifies the details of the scope, functionalities, and interactions of the software components. 
	Traditional software systems comprise of a finite number of states and the system behaviors are predictable. 
	However, for ML applications, the behavior of the program is unpredictable and defined by the training data. This makes the design of ML application a challenging task. 
	%
	%
	Again, ML applications require a large number of data. Thus, the design of ML applications need to accommodate the constraints and overhead on data processing.  
	As the algorithms and frameworks for machine learning are evolving rapidly, the ML application design should be flexible to adopt these changes. 
	
	Again, machine learning applications are data-driven and the performance of the system may degrade (\ie concept drift) over time despite no changes in the requirements or without the presence of bugs. 
	Thus, the maintenance requirement of the ML applications may be hard to predict. 
    So, the design need to be flexible to accommodate frequent changes. 
	%
	In cases of adding AI/ML capabilities to the existing application, the design should consider minimum restructuring of the existing system architecture. 
	Similarly, the design of a new ML application should be flexible to adapt future changes. 
	%
	In our case, we add functionalities to the existing application and 
	we focused on the functional requirements of the modules and the interfaces for interaction with the existing application. 
	
	\item[\textbf{S3:}] \textbf{\textit{Implementation}:}
	Development of ML applications commonly involves the use of many different frameworks and libraries \cite{Gartner_TPA_2017}. 
	However, it is hard to put together a diverse set of frameworks and libraries and to ensure compatibility and integrity of the system. 
	Again, ML models are ``Black Box" \cite{Honegger_MSc_2018} and are hardly explainable \cite{Doshi-Velez_arXiv_2018,Bibal_ESANN_2016}. Thus, it is hard to clearly understand why they work and why sometimes they do not. 
	In addition, ML models may exhibit robustness to  noises \cite{Grosse_and_Duvenaud_2014} making it harder to verify the implementation. 
	
	Again, the development environment for ML models might be different from the production environment. 
	So, the implementation of ML applications should consider the target platform requirements while choosing frameworks and libraries. Also, the hardware-software ecosystem for ML applications are rapidly evolving. Thus, the implementation choices should also consider maximizing portability, compatibility, and adaptability of ML applications with lower cost. 
	%
	In our project, we needed exploration and experimentation with different algorithms. 
	So, we implemented a framework that offers flexibility to apply different ML algorithms. Our implementation strategy leverages code reusability. We needed to use different ML frameworks (TensorFlow, Keras) to implement our ML-based components.  
	
	\item[\textbf{S4:}] \textbf{\textit{Integration}:}
	Software integration process aggregates the components of the system into one coherent unit with desired functionalities. Integration phase must ensure the functional integrity of the system by implementing appropriate interactions and coordination of the subsystems. The integration of ML applications can be viewed as a two-step process; first integrating the sub-components of the ML component and then the integration of the ML components with other non-ML components of the target system. Thus, the defined interface between ML components with other components may influence the process and complexity of the integration phase. Again, ML models are expected to evolve continuously. Thus, the ML workflow should facilitate continuous integration of ML models. 
	
	\item[\textbf{S5:}] \textbf{\textit{Testing}:} 
	As the outcomes of machine learning models are stochastic in nature \cite{Grosse_and_Duvenaud_2014, Murphy_SEKE_2007}, there might not be unique results to compare and verify machine learning applications. Thus, the existing unit testing frameworks (\eg PyUnit for Python) cannot be readily used to test ML applications. Again, ML models learn from the input data in an adaptive and iterative manner \cite{Pei_SOSP_2017}. The rules learned by the ML models depend on many parameters such as the selected features, the model architecture and even the training data. The rules generated by ML systems might even be unknown to the developers \cite{Pei_SOSP_2017}. Thus, it is harder to identify the erroneous system behaviors and to pinpoint the source of the bugs. 
	
	Again, ML algorithms may sometimes exhibit robustness to some bugs and produce reasonable outcomes by compensating for noisy data or implementation errors \cite{Grosse_and_Duvenaud_2014}. Thus, bugs in ML application can be tricky to detect and fix. 
	Testing of ML applications may involve large-scale training data and manual labeling of such data is costly. 
	Moreover, a random selection of subsets of data is likely to fail to identify many corner cases. All these issues make the testing and fixing of erroneous behavior in ML applications a very challenging task. 
	For our ML components,
	we evaluate the model accuracies with the evaluation data set. We also unit-tested individual modules and tested the overall functionality with integration testing. 
	
	\item[\textbf{S6:}] \textbf{\textit{Deployment}:} 
	Deployment phase puts the software system into production. Generally, the deployment phase either updates or replaces the existing system. 
	For machine learning applications, this phase is more likely to add new functional module(s) to the existing system. One important challenge to consider in ML system deployment is that the platform and infrastructure for production system might be very different from the environment the ML model was trained and evaluated \cite{Gartner_TPA_2017}. These differences can pose compatibility, portability and scalability challenges and may affect the performance. 
	
	In our project, we developed a framework to 
	deploy the models and to evaluate their performance. Our design aimed at creating an ML-based service for the existing applications. We tested our model deployment as API-based web services for error detection and correction. 
\end{itemize}
\vspace{-0.47cm}
\subsection{Machine Learning Perspectives} \label{secML}
ML application development has some specific aspects to consider as a software development project. The distinct characteristics and requirements of the ML application requires a well-defined set of principles and guidelines as recommended by practitioners \cite{Zinkevich_MLRules_2014}. 
\vspace{-0.1cm}
\begin{itemize}
	\item[\textbf{M1:}] \textbf{\textit{Problem Formulation}:} 
	Machine Learning algorithms offer general-purpose solutions. We need to formulate problems appropriately to fit into the ML-based solution space. 
	A wrong problem formulation may lead to failure of an ML application. 
	One key challenge to formulating a machine learning problem is that one should clearly understand the problem, the algorithms and the mapping of the problem from the original domain to the ML solution space. 
	The correct formulation of the problem is a prerequisite to the success of other phases of ML application development. 
	In this study, we analyzed the transaction processing workflow based on domain knowledge from industry experts. 
	Our problem definition is based on the domain understanding, transaction data structures and organization, types and characteristics of errors and their manual correction process. We also considered the relationships and dependencies among the transaction components. 
	
	\item[\textbf{M2:}] \textbf{\textit{Data acquisition}:} 
	Collection and processing of large volume of data are critical overheads for machine learning  \cite{Roh_data_survey_2018}. 
	Insufficient data is also a problem for machine learning applications. 
	The data acquisition must focus on the completeness (representative of full range of behaviours), accuracy (correctness of the data), consistency (no contradictory data), and timeliness (relevant to the current state of the system) of data to ensure data quality  \cite{Hebron_MLD_2016}. However, maintaining data quality requires careful steps in collection, curation, and maintenance of the data. This is often very expensive regarding the time and associated manual labour \cite{Hebron_MLD_2016}. 
	
	In our study, the transaction data was provided by SAP from a client company. 
	Sensitive business and personal information was removed from the transaction. 
	Although SAP solutions have a comprehensive set of features, the clients have the flexibility to customize data structures and functionalities. 
	These customization add challenges for the ML models to generalize for different clients for the same uses cases. 
	Our analysis suggests that some prior analysis of the data and the problem is important to set appropriate data requirements. 
	 
	\item[\textbf{M3:}] \textbf{\textit{Preprocessing}:}
	 Raw data may not be readily usable for machine learning models and may require different preprocessing for cleaning, organization, completion, and transformation. Noisy or dirty data is claimed to be the top challenge for ML practitioners \cite{Kaggle_survey_2017}. Noisy data can have adverse effects on the learning and thus inference of the models. Data preprocessing is an important and challenging phase for machine learning \cite{Sergio_survey_2017} and may incur a significant proportion (\textgreater 50\%) of time and effort  \cite{Herrera_multiple_2016}. 
	 In our study, we first gained an understanding of the attributes of the transaction data through careful analysis. 
	 We defined a set of constraints on the transaction data to filter data set. For example, we did not consider transactions that have more than 20 retail items. This constraint restricts features not to be too sparse while covering majority of the transactions. Our preprocessing steps also include filling the missing values and the  normalization of data fields to correct formatting differences.
	
	\item[\textbf{M4:}] \textbf{\textit{Feature Extraction}:}
	Feature extraction performs optimal transformation of input data into feature vectors for the machine learning algorithms \cite{Guyon_Springer_2008}. 
	Feature Extraction is very important for machine learning \cite{Storcheus_NIPS_2015}. 
	Feature extraction phase extracts a set of features that best represent the hidden characteristics of the data for machine learning. 
	Feature extraction also aims to remove the potential noises and redundancy from the data \cite{Guyon_Springer_2008}. This also finds a low-dimensional representation of the data to improve the speed of the training and inference of ML models \cite{Guyon_JMLR_2003}.  
	However, it is challenging to extract features by processing a high volume of data currently available. 
	Selection of features is important for model performance. 
	In our case, we considered the domain knowledge, the transaction processing workflow, and the transaction data structure and relationships as the basis of feature selection. 
	In addition to base features, we also extracted derived features for our ML models. 
	
	\item[\textbf{M5:}] \textbf{\textit{Model Building}:}
	ML models are created based on specific machine learning algorithms depending on the problem and the characteristics of the data. One may use existing models from the libraries or may create custom models. The models are then trained  iteratively until the models attain the desired level of performance. 
	One frequent problem for ML models is \textquoteleft overfitting\textquoteright. In this case, a model performs well on the training dataset but does not generalize. This might be because the model is too complex. The complexity of the model is related to higher dimensional features and the architecture of the model. The solution is to find the simplest model that does the required job. However, ML models should not be too simple as they would likely fail to capture the hidden patterns in the data causing \textquoteleft underfitting\textquoteright. 
	Also, data distribution need to be balanced. Otherwise, the inference of the model might be biased towards the dominating class of the training data. 
	In our project, we observed overfitting issues when we used neural networks of comparatively deeper architecture for performance gain in error classification. We eliminated feature redundancies and tuned neural network architectures to avoid overfitting. Class-imbalance problem was another important issue in our study as the distribution of error samples were very skewed to a limited class of errors. However, we prepared the training, test, and evaluation dataset as balanced as possible to improve model performance. 
	 \item[\textbf{M6:}] \textbf{\textit{Evaluation}:}
	ML models are evaluated by applying the ML models to the evaluation dataset exclusively separated from the training data set. 
	Both pre-deployment evaluation and post-deployment performance monitoring are important as the performance may drop with changes in the input data characteristics over time. 
	Thus, ML models may need to update (\eg retrain) to adapt to the changes.
	So, the evaluation of ML models can be iterative in the model life cycle.
	Also, there might be different interacting models. 
	So, the performance of the individual models may only represent a part of the end-to-end scenarios. 
	Thus, it is important to have both model-level and system-level evaluations. 
	In our study, we evaluated the individual models with the evaluation data set. 
	We also evaluated the overall performance after the integration of the models. 
	 
	\item[\textbf{M7:}] \textbf{\textit{Model integration and deployment}:}
	Once trained models are available, the next step is to integrate the models into the target application. 
	This involves putting all necessary components (e.g., models, input-output pipelines) together. 
	For multiple models, it may require to define and implement the interface for each model to interact with other models and components of the target application. One common approach of deploying ML models is to deploy them as services and accessing the services through APIs. 
	The deployment of ML models should consider the portability and compatibility of the models regarding the target platform. 
	In our project, integration and deployment involve developing an interface between the ML-based components and the existing application to provide the error detection and correction services. 
	
	\item[\textbf{M8:}] \textbf{\textit{Model management}:} 
	ML model management involving training, maintenance, deployment, monitoring, and documentation of ML models is a challenging task in ML workflow \cite{Schelter_IEEE_bulletin_2018}. 
	ML models are data-driven and are based on different assumptions on the distributions and patterns of data. 
	However, initial characteristics of data may not hold due to changes in the data. This can also affect model behaviour. 
	Thus, it is important to monitor the performance of the deployed models, track changes in the data characteristics, and also, retrain and re-validate the models. 
	These require iterations on ML model life-cycle activities 
	which are often very expensive regarding time and resources. 
	
	Again, ML models involve multifaceted variabilities such as configurations, model parameters, dataset, etc. 
	However, it is important but hard to keep track of the model versions, dataset and configurations to allow \textit{reproducibility} of ML models \cite{Hermann_Uber_2017, Zaharia_Databricks_2018} and easier management of ML workflow \cite{Jeffrey_FBookAI_2016}. 
	Model reproducibility helps us to analyze and compare model behaviour and performance, and also supports deployment or roll out decisions. 
	We defined policies to version data, keep track of models and configurations for comparative analysis of our ML models. 
	\vspace{0.5cm}
	\item[\textbf{M9:}] \textbf{\textit{Ethics in AI development}:} 
	Machine learning is advancing rapidly to influence an increasing number of areas through software and services. 
	It is very important to ensure that the use AI or ML conforms with the ethical standard to avoid any negative consequences. 
	Researchers and practitioners should perform 
	\textquotedblleft responsible\textquotedblright~use of AI \cite{Google_responsibleAI}. 
	The teams involved in the research and development of an ML application should adhere to the standard code of ethics for Software Engineering \cite{SE_code_of_ethics_1999}. 
	In our project, we maintained strict data privacy and security policies. Personal information was filtered out from the transactions and all the team members followed a non-disclosure agreement that protects the privacy and security of personal and business-sensitive information. 
\end{itemize}
\vspace{-0.3cm}
\subsection{Research Collaboration Perspectives} \label{secCollab}
Effective collaboration is a key requirement for the success of a collaborative project. 
Collaboration must aim to bring the best out of the partnering teams towards the research and development goals. 
However, multi-partners (more specifically industry-academia) collaboration adds some inherent challenges \cite{Sandberg_ICSE_SEIP_2017} in terms of communication, interactions, and common understanding of the problem. 
We shade light on some collaboration challenges below: 
\begin{itemize}
	\item [\textbf{C1:}] \textbf{\textit{Problem Understanding}:} 
	 In a collaborative project, it is important for the partners to have the same or closely similar understanding of the problem. 
	 Collaborations may be based on an academic research idea or a problem from the industry. 
	 The collaborative teams must understand the problem clearly to translate it into an appropriate research problem. 
	 The research aspects of the problem might be more transparent to the academic team(s) while the industry partner(s) may share domain knowledge and the business cases of the problem.  
	 
	\item [\textbf{C2:}] \textbf{\textit{Focus on objectives}:} 
	In the industry-academia collaborative project, there might be differences in prioritizing the objectives by individual partners. For example, the academic collaborators may naturally focus more on the research outcomes and may be more inclined to measure success in terms of research findings. 
	Industry partner, on the other hand, might be more focused on maximizing the business values from the investment of time and resources. 
	As the differences in objectives are not mutually exclusive, the differences in priorities may impact positively on the overall outcomes of the project. Thus, the heterogeneity in partnerships is more likely to enforce a balance between focuses on research and development. 
	This, in turn, will translate the research outcomes into more practically usable deliverables as product and services to add business values.   
	     
	\item [\textbf{C3:}] \textbf{\textit{Knowledge Transfer}:} 
	Knowledge transfer is an important driver of innovation and economic growth in industry-academia collaborations \cite{Vries_JTT_2018, Alexander_PPC_2013}. 
	In industry-academia collaboration on software engineering research, one of the key objectives is to transfer the research outcomes (\eg knowledge and technology) from the academic partner(s) to the industry partner(s) and vice-versa. 
	Such collaboration bridges the gap between research and practice. 
	However, studies show that technological relatedness and technological capability are important facilitators for knowledge transfer while tacitness and ambiguity may affect knowledge transfer \cite{Santoro_TEM_2006}. 
	Also, the differences in organizational cultures may influence the interpretation of behaviour, interactions, and knowledge. 
	 
	\item [\textbf{C4:}] \textbf{\textit{Professional Practice}:} 
	In an industry-academia collaboration, the heterogeneity of team members with research and professional software development background is an asset. The blend of research and development skills and the domain expertise add diversity in overall team skill set. However, professional practices and work process are expected to be different in academia and industry. The workplace practices might be different due to the nature of the day-to-day work. In industry, the work process is more formal and might be influenced by corporate business cultures. While in academia, the work process is more outcome-focused with a comparatively flexible or less formal process. Many studies acknowledged these cultural differences relevant to the industry-academia collaboration \cite{Bruneel_RP_2010, Galan-Muros_RDM_2016}. 

	\item [\textbf{C5:}] \textbf{\textit{Data Privacy and Security}:} 
	Data privacy and security is crucial when business data with sensitive user information is subject to analysis for research and development. Application of machine learning may require access to a large volume of such data. Thus, the privacy and security of data are highly important. In addition to ensuring the privacy and security, use of data for AI must comply with ethical standards (see M8). Usually, data privacy and security is governed by the mutual agreement among the partnering organizations. The team members need to be trained on the privacy and security of the data and associated resources (e.g., devices, networks). 
\end{itemize}

\vspace{-0.2cm}
\subsection{Recommendations}
Researchers and practitioners in software engineering and machine learning are in need of a consolidated set of guidelines and recommendations for research and development of ML applications. We summarize our guidelines or recommendations from the perspectives of our three identified dimensions of challenges for ML applications.  
\subsubsection{\textbf{Software Engineering Perspective}} 
From the software engineering perspective, we recommend the following guidelines to address the challenges we identified in Section \ref{secSE}. 
\begin{itemize}
    \item \textit{Requirements Engineering}
        \begin{itemize}
            \item ML models are data-driven and thus it is important to analyze the feasibility of whether the available data are suitable for intended ML-based solutions.  
            
            \item Applications can have requirements of both ML-specific and non-ML types. Developers should be aware of any conflicting requirements and adapt the requirements accordingly. 
            
            \item Direct communication with the end users is important to identify requirements precisely. A late discovery of the missed requirements can be costly to accommodate. 
            
            \item 
            Requirements in ML applications can evolve frequently. Thus, it is important to iteratively refine requirements. 
        \end{itemize}
        \vspace{0.25cm}
    \item \textit{Design}
        \begin{itemize}
            \item ML applications are expected to be modular in design as modularity offers separation of concerns and reusability. Each component can be developed with cohesive functionalities. 
            The overall system is built by the integration of interacting modules. 
            \item ML components may evolve faster than other system components due to frequent changes in requirements and data. The design thus should be flexible to accommodate changes with minimum effort and cost. The design of ML components needs to be lightly coupled to avoid possible maintenance cost.  
            \item ML applications require a large volume of data. Thus, the design must accommodate the appropriate data handling mechanism in the system.  
            \item As the models or components need to be integrated through appropriate interfaces, integration strategies must be reflected in the design. 
        \end{itemize}
    \item Implementation
        \begin{itemize}
            \item Implementation of ML applications should aim to use a cohesive set of frameworks and libraries. 
            \item Implementation should consider the target platform of the application for scalability, compatibility, and portability. 
        \end{itemize}
        \vspace{0.5cm}
    \item Integration
        \begin{itemize}
            \item Integration aggregates the system components. This must ensure the functional integrity of the system.
            \item Integration process must reflect the design considerations specified in the design of the system. 
        \end{itemize}
    \item Testing
        \begin{itemize}
            \item ML models are opaque and it is hard to explain the model behaviours. 
            ML models need to be tested rigorously in a wide variety of settings and for all possible range of use-case scenarios. 
            \item Bugs in ML application are hard to detect because of the stochastic nature of ML systems. Both ML and non-ML components should be carefully unit tested as well as their integration. 
        \end{itemize}
    \item Deployment
        \begin{itemize}
            \item The deployment should consider the platform-specific differences in development and production environment.
            \item Deployment of ML applications in a production environment must consider portability and scalability requirements. 
            \item 
            The deployment and roll-out strategy should be cautious not to affect the system users. 
        \end{itemize}
\end{itemize}

\vspace{-0.65cm}
\subsubsection{\textbf{Machine Learning Perspective}}  
From the machine learning (ML) perspective, we recommend the following guidelines to address the challenges we identified in Section \ref{secML}.
\vspace{-0.2cm}
\begin{itemize}
    \item Problem Formulation 
        \begin{itemize}
            \item The correct formulation of a problem as an ML problem is a prerequisite for the success of ML applications. Formulation of ML problems must be based on a clear understanding of the target  
            problem and the characteristics of data. 
            \item ML algorithms offer general-purpose solutions. So, developers must understand the algorithms to select appropriate algorithm(s) for a particular problem. 
        \end{itemize}
        \vspace{0.05cm}
    \item Data Acquisition
        \begin{itemize}
        \item Data collection process should ensure the completeness, accuracy, consistency, and timeliness of the dataset. 
        \item The structure of the data may evolve over time, the data acquisition process should be flexible to easily adapt changes in data structure and organization. 
        \item Data requirements should be set by necessary analysis prior to the acquisition of a large volume of data. This may save considerable time and resources in the ML workflow. 
        \end{itemize}
    \item Data Preprocessing
        \begin{itemize}
            \item For data-driven system, the fact is ``garbage in garbage out''. Thus, raw data need preprocessing to remove noise, to fill in the missing values, and to do other transformations. 
            \item Required preprocessing might be common to modules. Thus, the overall workflow should maximize the reusability of the preprocesed data.            
        \end{itemize}
    \item Feature Extraction
        \begin{itemize}
            \item The quality of features greatly influences the performance of the ML models. Thus, the developers must find the best set of features. 
            \item An automated pipeline should be built for feature extraction as it is time and resource consuming. 
            \item Noisy features can have adverse effects on the model performance. So, noises need to be cleaned from features. 
            \item The dimension of the features is related to the complexity of the models. So, features need to be represented in a possible lower dimension but preserving hidden characteristics. 
            \item There should be a periodic review of the features \cite{Zinkevich_MLRules_2014}.
        \end{itemize}
    \item Model Building
        \begin{itemize}
            \item ML algorithms need to be tailored based on the problem and the characteristics of the data. 
            \item 
            ML developers should start with a simpler solution and gradually adopt more complex solutions considering the resource-performance trade-off.
            \item 
            The developers must ensure quality and balanced distribution of data.  
            \item Whenever it is possible, developers should consider reusing existing solutions (for productivity) before opting for customized or more innovative solutions. 
        \end{itemize}
    \item Model Evaluation
        \begin{itemize}
         \item 
         The evaluation dataset should be complete enough to represent all possible use-case scenarios. 
         \item Model evaluation should focus on precision/recall/accuracy but also on other factors such as throughput, resource utilization, and scalability. 
         \item When different models interact, both model-level and system-level performances need to be evaluated.
        \end{itemize}
    \item Model Integration and Deployment
        \begin{itemize}
            \item Integration of ML components must ensure the functional integrity of the ML applications. 
            \item The deployment should consider the compatibility and differences in development and production platforms.
            \item The deployment must ensure a smooth roll-out of the existing system without affecting the user or the business process.  
        \end{itemize}
        \vspace{0.15cm}
    \item Model Management
        \begin{itemize}
        \item Post-deployment monitoring of models is important. 
        Based on the performance, models may need to be retrained which may initiate a maintenance cycle. 
        \item Data need to be monitored because the data characteristics and distribution may change over time.  
        \item 
        Building a model with desired performance requires exploration and experimentation. 
        Versioning of the models, data, and configurations is important to facilitate reproducibility of ML models. 
        \item Each phase of the ML life-cycle should be well-documented to support model maintenance. 
        \end{itemize}
    \item Ethics in AI development
        \begin{itemize}
            \item Development of ML applications must adhere to AI ethics principles to ensure responsible use of AI.
            \item Privacy and security of personal and business data must be preserved. 
            \item Collective well-being must get priority over business gains in the use of AI and ML.
        \end{itemize}
\end{itemize}
\vspace{-0.3cm}
\subsubsection{\textbf{Industry-academia Collaboration Perspective}}
We recommend the following guidelines to address the challenges from the industry-academia collaboration perspective presented in Section \ref{secCollab}:
\vspace{-0.15cm}
\begin{itemize}
    \item Problem Understanding 
        \begin{itemize}
            \item All members of the team should have the same understanding of the problem. 
            Sharing of perceptions can be useful to understand the problem from diverse perspectives. 
        \end{itemize}
     \item Focus on Objectives
        \begin{itemize}
            \item The success of the collaboration largely depend on the collective focus of the teams. The focus should thus be on the commonalities of the objectives than differences.
            \item Objectives need to be prioritized with consensus. 
            Differences in objectives should be addressed in a cooperative manner than a competitive manner. 
        \end{itemize}
        \vspace{0.5cm}
    \item Knowledge Transfer
        \begin{itemize}
            \item Knowledge and technology transfer is a primary objective of industry-academia collaboration. 
            Frequent interactions can be useful in easing knowledge transfer. 
            \item Identification of the differences in knowledge and professional backgrounds can ease knowledge transfer.
            \item Training on academic and industrial perspectives may bridge the knowledge gaps among the partners.  
        \end{itemize}
    \item Professional Practice
        \begin{itemize}
            \item It is important to reach a common ground despite the differences in professional practices and institutional cultures to collectively focus on the objectives. 
            \item Interactions and collaborations can be useful  to bridge the gap in professional cultures.
        \end{itemize}


    \item Data Privacy and Security
        \begin{itemize}
            \item Privacy and security of sensitive personal and business data must be ensured.
            \item Policies must be defined and communicated to the teams to ensure the privacy and security of sensitive data.
        \end{itemize}
\end{itemize}

\vspace{-0.5cm}
\section{Related Works} \label{lblRelatedWorks}
Recent advancements in machine learning algorithms, frameworks, and  high-performance computing, and availability of large volume of data are making ML increasingly popular to devise innovative solutions for diverse problems. 
However, increasing adoption of machine learning into software applications is posing additional challenges to software development process. 
Challenges in traditional software engineering process have been widely addressed by the researchers \cite{Sandberg_ICSE_SEIP_2017, Garousi_IST_2017}. 
However, there is a growing need for guidelines and best practices for developing ML applications especially in collaborative research and development context. 

\vspace{0.1cm}
Sandberg and Crnkovic \cite{Sandberg_ICSE_SEIP_2017} outlined important challenges in industry-academia collaboration and   
reported that the agile methodology (\eg SCRUM) fits well into such collaboration. 
Garousi \etal \cite{Garousi_IST_2017}, synthesized the challenges and best practices, and recommended guidelines for industry-academia collaborations in software engineering through an agile approach. 
Ankrah and AL-Tabbaa \cite{Ankrah_SJM_2015} also presented some key aspects of industry-academia collaborations. 
Schelter \etal \cite{Schelter_IEEE_bulletin_2018} focused on ML model management regarding use cases from conceptual, data management, and engineering perspectives. 
Amershi \etal \cite{Amershi_ICSE_2019} highlighted challenges in AI application development in Microsoft and shared how the teams address those challenges. 
Zinkevich \cite{Zinkevich_MLRules_2014} presented a concise set of guidelines for best practices in ML engineering. 
However, these guidelines are not focused to find a fit for ML into traditional software engineering process. 
Sapp \cite{Gartner_TPA_2017} provides some professional advice on the architecture and organization of ML projects. 
Jordan \cite{Jordan_ML_project_2018} presents a set of guidelines for machine learning project management. 
Many existing literatures focus on different aspects of machine learning such as data acquisition \cite{Roh_data_survey_2018,Attenberg_selective_2014}, data preprocessing \cite{Sergio_survey_2017,Hinton_Dimensionality_2008}, feature extraction \cite{Guyon_Springer_2008,Storcheus_NIPS_2015}, model management \cite{Schelter_IEEE_bulletin_2018}, testing \cite{Pei_SOSP_2017,Grosse_and_Duvenaud_2014} and deployment \cite{Schelter_IEEE_bulletin_2018} of ML applications. 
Besides, several studies focus on the challenges in traditional software engineering \cite{Attarha_ICIS_2011,Firesmith_JOT_2007}. 

\vspace{0.1cm}
Existing literature shared useful insights on challenges for machine learning and software engineering separately. 
However, there is a growing need for a consolidated set of guidelines regarding software engineering for machine learning. This is especially important when collaborations of diverse teams from industry and academia are involved. 
We aim to contribute to filling this important gap by reporting about our experience building ML software components in an industrial context. 
\vspace{-0.25cm}
\section{Conclusion} \label{lblConclusion}
Artificial intelligence and machine learning are being increasingly adopted in modern software applications. With diverse benefits, ML is also adding new challenges in software development process. 
In this paper, we share our experience on the development of ML-based automatic detection and correction of errors in retail transactions. We outline our detail approach involving research and development to solve this important real-world problem in the retail business. We identify several dimensions of challenges in ML application development, especially in an industry-academia collaboration context. Based on our experience building these ML components and the principles and practices in software engineering and machine learning, we report some important insights and highlight some recommendations that we believe will be useful to researchers and practitioners embarking in engineering ML applications. 

\end{document}